\documentclass{IEEEtran}
\def\BibTeX{{\rm B\kern-.05em{\sc i\kern-.025em b}\kern-.08em
		    T\kern-.1667em\lower.7ex\hbox{E}\kern-.125emX}}

\let\labelindent\relax	% required for using enumitem

\usepackage{amsmath,amssymb,amsfonts}
\usepackage{mathabx}
\usepackage{algorithmic}
\usepackage{graphicx}
\usepackage{textcomp}
\usepackage[utf8]{inputenc}
\usepackage[T1]{fontenc}
\usepackage{amsmath}
\usepackage{amsfonts}
\usepackage{amssymb}
\usepackage{balance}
\usepackage{graphicx}
\usepackage[rightcaption]{sidecap}
\usepackage{import}
\usepackage{textcomp}
\usepackage[dvipsnames]{xcolor}
\usepackage{bbm}
\usepackage{slashed}
\usepackage{caption}
\usepackage{subcaption}
\usepackage{lipsum}
\usepackage{enumitem}
\usepackage{tabularx}
\usepackage{xcolor}
\usepackage[ruled,linesnumbered]{algorithm2e}
\usepackage{leftindex}
\usepackage{mathtools}
\usepackage{booktabs}
\usepackage{makecell}
\usepackage[noadjust]{cite}
\usepackage{multirow}
\usepackage{url}

\usepackage{easyReview}
\usepackage{bm}

\SetKwComment{Comment}{/* }{ */}
\SetKwInOut{Input}{Input}
\SetKwInOut{Output}{Output}
\SetKwInOut{Parameters}{Parameters}

\newcommand{\bbR}{\mathbb{R}}

\newcommand{\calC}{\mathcal{C}}
\newcommand{\calD}{\mathcal{D}}

\newcommand{\calH}{\mathcal{H}}

\newcommand{\calK}{\mathcal{K}}

\newcommand{\calO}{\mathcal{O}}
\newcommand{\calP}{\mathcal{P}}

\newcommand{\calU}{\mathcal{U}}

\newcommand{\calX}{\mathcal{X}}

\usepackage{amsthm}

\theoremstyle{definition}
\newtheorem{assumption}{Assumption}%[section]
\newtheorem{theorem}{Theorem}
\newtheorem{lemma}[theorem]{Lemma}
\newtheorem{proposition}[theorem]{Proposition}
 
\newtheorem{definition}{Definition}%[section]

\theoremstyle{remark}
\newtheorem{remark}{Remark}

\DeclareMathOperator*{\argmin}{arg\,min}

\begin{document}
\title{Decoupled Design of Time-Varying Control Barrier Functions via Equivariances}
\author{Adrian Wiltz and Dimos V. Dimarogonas
\thanks{This work was supported by the ERC Consolidator Grant LEAFHOUND, the Swedish Research Council, Swedish Foundation for Strategic Research, and the Knut and Alice Wallenberg Foundation.}
\thanks{The authors are with the Division of Decision and Control Systems, KTH Royal Institute of Technology, SE-100 44 Stockholm, Sweden {\tt\small \{wiltz,dimos\}@kth.se}.}}

\maketitle
\begin{abstract}
	This article presents a systematic method for designing time-varying Control Barrier Functions (CBF) composed of a time-invariant component and multiple time-dependent components, leveraging structural properties of the system dynamics. The method involves the construction of a specific class of time-invariant CBFs that encode the system’s dynamic capabilities with respect to a given constraint, and augments them subsequently with appropriately designed time-dependent transformations. While transformations uniformly varying the time-invariant CBF can be applied to arbitrary systems, transformations exploiting structural properties in the dynamics --- equivariances in particular --- enable the handling of a broader and more expressive class of time-varying constraints. The article shows how to leverage such properties in the design of time-varying CBFs. The proposed method decouples the design of time variations from the computationally expensive construction of the underlying CBFs, thereby providing a computationally attractive method to the design of time-varying CBFs. The method accounts for input constraints and under-actuation, and requires only qualitative knowledge on the time-variation of the constraints making it suitable to the application in uncertain environments. 
\end{abstract}

\begin{IEEEkeywords}
Control Barrier Functions, Constrained Control, Time-Varying Systems, Symmetry in Control Systems, Safety-Critical Control.
\end{IEEEkeywords}

%%%%%%%%%%%%%%%%%%%%%%%%%%%%%%%%%%%%%%%%%%%%%%%%%%%%%%%%%%%%%%%%%%%%%%%%%%%%%%%%
% INTRODUCTION

\section{Introduction}

Safety-critical systems commonly evolve in time-varying environments, which are upon controller design not or only partially known. This poses significant challenges to the design of controllers ensuring constraint satisfaction, typically formalized as the problem of rendering a subset of the state space forward invariant. The controller design becomes especially challenging for systems with degenerate controllability properties such as for systems subject to input constraints or due to kinematics (e.g. non-holonomic constraints). 

As a control theoretic concept, Control Barrier Functions (CBF)~\cite{Wieland2007,Ames2017} characterize forward control-invariant sets with respect to state constraints for general dynamics over infinite time-horizons. While the synthesis of CBFs is challenging, substantial progress has been made in this respect, and systematic design approaches have been proposed based on optimization (SOS, predictive methods, reachability, etc.), data-driven concepts or analytic construction. Most of these works, however, focus on time-invariant constraints. Yet, time-varying constraints arise in many applications, including the avoidance of time-varying obstacles~\cite{Huang2023,Haraldsen2024a}, human-robot collaboration~\cite{FernandezAyala2023,Shi2024}, safely landing a UAV on a moving platform~\cite{Sankaranarayanan2024,Persson2019}, or handling high-level tasks through forward invariant sets~\cite{Lindemann2019,Wang2023a}. %Also the coordination of vehicles through state-coupled CBFs gives rise to constraints varying over time~\cite{Frauenfelder2023}. 
% new paragraph can be inserted in extended version
The handling of time-varying constraints in a general setting is more challenging than in the time-invariant case and formal guarantees are more difficult to establish, especially when facing actuation limitations. The approach of augmenting the state space by time as another state becomes quickly intractable, especially for constraints defined on infinitely long time horizons. Moreover, such an approach requires full knowledge on the time-variation of the constraints a priori and changes in the constraint specification necessitate the recomputation of the CBF.

We mitigate these problems by decoupling the synthesis of time-varying CBFs into two parts: (1) the synthesis of a time-invariant, only state-dependent CBF, and (2) the design of a time-dependent transformation. While~\cite{Wiltz2024a} shows that such an approach is universally viable for the synthesis of uniformly time-varying CBFs, taking the form 
\begin{align*}
	B_{\text{uniformly}}(t,x) \coloneq b(x) + \bm{\lambda}(t),
\end{align*}
we show in this article how such an approach can be extended to more general time-variations for systems exhibiting equivariances. Equivariances~\cite{Field1980,Mahony2020} can be, loosely speaking, viewed as some type of symmetry in the dynamics and have been previously employed in the synthesis of time-invariant CBFs~\cite{Bousias2025,Wiltz2025c}. 

In this article, we provide conditions under which the CBF properties of a given CBF~$ b $ are preserved by transformations taking the form
\begin{align*}
	B(t,x) \coloneq b(D(x;t)) + \bm{\lambda}(t),
\end{align*}
where $ D: \bbR^{n}\times\bbR_{\geq0} \rightarrow \bbR^{n} $ is a time-parameterized transformation to be further specified below, and show that $ B $ constitutes a CBF with respect to the time-augmented dynamics. The proposed method only requires qualitative knowledge of the time-variations of the constraints in terms of an upper-bound on the rate of change, making it suitable for the application in uncertain environments. Moreover, the method explicitly accounts for input constraints and under-actuation.

A related approach in~\cite{Braun2024} shifts Lyapunov functions along a predefined trajectory, handling constraints by adapting the progression rate to keep the system state within a sufficiently small sublevel set of the CLF. In contrast, our approach identifies the maximum allowable rate of change in advance, enabling its application to time-varying constraints. Methods based on navigation functions or high-gain controllers~\cite{Koditschek1990} can also address time-varying constraints, but handling systems with degenerate controllability properties or input constraints is challenging if at all possible. 

\emph{Notation:} Let $ \calX\subseteq\bbR^{n} $, $ x\in\bbR^{n} $. We denote sets by calligraphic upper case letters, while trajectories $ \bm{x}:\bbR \rightarrow \calX $ are denoted by boldface lower case letters. The composition of $ f:\bbR^{l}\rightarrow\bbR^{m} $ and $ g:\bbR^{m}\rightarrow\bbR^{n} $ is $ g\circ f: \bbR^{l}\rightarrow\bbR^{n} $. The directional derivative (Dini derivative) in the direction of $ v_1, v_{2}, $ etc. of a locally Lipschitz continuous function $ \phi: \bbR^{m_{1}}\times\bbR^{m_{2}}\times\dots \rightarrow \bbR^{n} $ is $ df(x_{1}, x_{2},\dots;v_1,v_2,\dots) = \liminf_{\varepsilon\downarrow 0} \frac{f(x_{1}+\varepsilon v_{1}, x_{2}+\varepsilon v_{2}, \dots) - f(x_{1},x_{2},\dots)}{\varepsilon} $.

%%%%%%%%%%%%%%%%%%%%%%%%%%%%%%%%%%%%%%%%%%%%%%%%%%%%%%%%%%%%%%%%%%%%%%%%%%%%%%%%
% PRELIMINARIES

\section{Background}
\label{0008_sec:background}

Consider the input-constrained system
\begin{align}
	\label{0008_eq:dynamics}
	\dot{x} = f(x,u), \qquad x(0) = x_{0}
\end{align}
with $ x,x_{0} \in \bbR^{n} $, $ u\in\calU\subseteq\bbR^{m} $, and $ f: \bbR^{n}\times\calU\rightarrow\bbR^{n} $ being Lipschitz continuous in both of its arguments ensuring existence and uniqueness of the system's solution at least locally; forward completeness is assumed. 

\subsection{Equivariances}
Of particular interest in this article are dynamics that exhibit equivariances, which allow to generalize the local behavior of a dynamic system to a larger set of states. 

\begin{definition}(Equivariant Dynamics)
	\label{0008_def:equiv}
	Consider a non-trivial diffeomorphism $ D_{f}(\cdot;p): \bbR^{n}\rightarrow\bbR^{n} $ dependent on an (optional) parameter $ p\in\calP\subseteq\bbR^{n} $. Dynamics~\eqref{0008_eq:dynamics} are called \emph{equivariant with respect to $ D_{f} $ and $ \calP $} if there exists some (possibly trivial) isomorphism $ D_{u}(\cdot;p): \calU \rightarrow \calU $, with $ D_{u}(\calU;p)\subseteq\calU $ for all $ p\in\calP $ such that 
	\begin{align}
		\label{0008_eq:def equivariance} 
		f(D_{f}(x;p),D_{u}(u;p)) = J_{D_{f}}(x;p) f(x,u) \qquad \forall p\in\calP, 
	\end{align}
	where $ J_{D_{f}} $ denotes the Jacobian to $ D_{f} $ with respect to its first argument.
\end{definition}

Thus, an equivariant system exhibits in states $ x $ and $ x'=D(x;p) $ for any $ p\in\calP $ the same dynamic behavior. We call an isomorphism $ D:\bbR^{n}\rightarrow\bbR^{n} $ trivial if it maps each point to itself, that is, $ D(x)\equiv x $; it is \emph{non-trivial} if it is not trivial. 

\subsection{CBFs in the Dini sense}
For the sake of generality, we consider CBFs in the Dini sense, which are only required to be differentiable almost everywhere except on a set of measure zero. 

\begin{definition}[CBF in the Dini sense~\cite{Wiltz2025b}]
	\label{0008_def:cbf}
	Consider $ \calD\subseteq\bbR^{n} $ and a locally Lipschitz continuous function $ b:\bbR^{n}\rightarrow\bbR $ such that $ \calC\coloneq\{ x\in\bbR^{n} \, | \, b(x)\geq0 \} $ is compact and $ \calC\subseteq\calD\subseteq\bbR^{n} $. We call such $ b $ a \emph{CBF in the Dini sense} on $ \calD $ with respect to~\eqref{0008_eq:dynamics} if there exists an extended class~$ \calK_{e} $ function $ \alpha $ such that for all $ x\in\calD $ 
	\begin{align}
		\label{0008_eq:cbf condition}
		\sup_{u\in\calU} \{db(x;f(x,u))\} \geq -\alpha(b(x)),
	\end{align}  
	where $ d\phi(x;v) $ with $ \phi:\bbR^{n} \rightarrow \bbR $ locally Lipschitz continuous denotes the \emph{Dini derivative at $ x $ in direction $ v $} as $ d\phi(x;v)\coloneq\liminf_{\varepsilon\downarrow0} \tfrac{\phi(x+\varepsilon v) - \phi(x)}{\varepsilon} $. 
\end{definition}

The previous definition includes the definition of differentiable CBFs~\cite{Ames2017} as a special case, and is analogous to the formulation of CLFs in the Dini sense~\cite{Clarke2011}. In the sequel, we always refer to a CBF in the Dini sense when writing CBF. For the construction of time-varying CBFs, we focus on a subclass of CBFs.

\begin{definition}[Shiftable CBF~\cite{Wiltz2024a}]
	\label{0008_def:shiftable cbf}
	A Lipschitz continuous function $ b:\bbR^{n}\rightarrow\bbR_{\geq0} $ is called a \emph{$ \Lambda $-shiftable CBF} with respect to~\eqref{0008_eq:dynamics} for some $ \Lambda>0 $ if $ b(x) $ is a CBF on the domain $ \calC_{\Lambda} \coloneq \{x\, | \, b(x)\geq -\Lambda\} $ with respect to~\eqref{0008_eq:dynamics}, or equivalently, if there exists an extended class~$ \calK_{e} $ function $ \alpha $ such that~\eqref{0008_eq:cbf condition} holds for all $ x\in\calC_{\Lambda} $. 
\end{definition} 

Such function is called shiftable as $ b(x)+\lambda $ constitutes a CBF for any $ \lambda\in[0,\Lambda] $ as well. We denote its super-level sets by
\begin{align*}
	\calC_{\lambda}\coloneq \{ x \, | \, b(x)\geq-\lambda\}, \qquad \lambda\in[0,\Lambda].
\end{align*}
Under certain conditions, $ \lambda $ can be even selected as a time-varying function while the property of the overall function as a CBF is preserved~\cite{Wiltz2024a}. Finally, we define time-varying CBFs. To this end, it suffices to assume relaxed continuity properties with respect to time, without impacting the forward control-invariance guarantees~\cite{Wiltz2025d}.

\begin{assumption}
	\label{0008_ass:relaxed continuity for tv cbf}
	Let $ B: \bbR_{\geq 0}\times\bbR^{n}\rightarrow\bbR $ be piecewise differentiable and upper semi-continuous with respect to its time argument, and it holds
	\begin{align*}
		\lim_{t\uparrow t_{0}} B(t,x) \leq \lim_{t\downarrow t_{0}} B(t,x) \qquad \forall x\in\bbR^{n}, \; \forall t_{0}\geq 0.
	\end{align*}
\end{assumption}

\begin{definition}[Time-Varying CBF \cite{Wiltz2025d}]
	\label{0008_def:tv cbf}
	Consider $ \calD\subseteq\bbR^{n} $ and a function $ B $ satisfying Assumption~\ref{0008_ass:relaxed continuity for tv cbf} such that $ \calC(t)\coloneq\{x\,|\, B(t,x)\geq 0\} $ is compact and $ \calC(t)\subseteq\calD\subseteq\bbR^{n} $ for all $ t\geq0 $. We call such $ B $ a \emph{time-varying CBF in the Dini sense} on domain $ \calD\subseteq\bbR^{n} $ with respect to~\eqref{0008_eq:dynamics} if there exists an extended class~$ \calK_{e} $ function $ \alpha $ such that for all $ (t,x)\in\bbR_{\geq0}\times\calD $
	\begin{align}
		\label{0008_eq:time-varying cbf condition}
		\sup_{u\in\calU} \{dB(t,x;1,f(x,u))\} \geq -\alpha(B(t,x)),
	\end{align}
	where $ dB(t,x;1,v) \coloneq \liminf_{\varepsilon\downarrow0} \tfrac{B(t+\varepsilon, x+\varepsilon v) - B(t,x)}{\varepsilon} $ denotes the Dini derivative at $ (t,x) $ in direction $ (1,v) $.
\end{definition}

The forward invariance results coming along with CBFs~\cite{Ames2017} stay analogous for time-varying CBFs in the Dini sense.

\begin{proposition}[\cite{Wiltz2025d}]
	Let $ B $ be a time-varying CBF in the Dini sense with respect to~\eqref{0008_eq:dynamics} on $ \calD\subseteq\bbR^{n} $. Moreover, let $ \bm{u}:\bbR_{\geq0} \rightarrow\calU $ be piecewise continuous and the corresponding state trajectory $ \bm{\varphi}(\cdot;x_{0},\bm{u}) $ starting in some initial state $ x_{0}\in\calC(0) $. If $ dB(t,x_{t};1,f(x_{t},u_{t})) \geq -\alpha(B(t,x_{t})) $ for all $ t\geq 0$, where $ x_{t}\coloneq\bm{\varphi}(t;x_{0},u) $ and $ u_{t}\coloneq\bm{u}(t) $, then $ \calC(\cdot) $ is forward invariant such that $ x_{t}\in\calC(t) $ for all $ t\geq 0 $.
\end{proposition}

\subsection{Auxiliary Result on Comparision Functions}

A function $ \alpha:\bbR_{\geq0}\rightarrow\bbR $ is called a class~$ \calK $ function if it is strictly increasing and $ \alpha(0) = 0 $; it is called an extended class~$ \calK_{e} $ function if it is extended to $ \alpha: \bbR \rightarrow \bbR $~\cite{Khalil2002}. The following result upper-bounds the sum of a class~$ \calK $ and an extended class~$ \calK_{e} $ function by another extended class~$ \calK_{e} $ function. 

\begin{lemma}[\cite{Wiltz2025d}]
	\label{0008_lemma:upper bound class K function}
	Let $ \alpha_{1} $ be an extended class~$ \calK_{e} $ function and $ \alpha_{2} $ a convex or concave class~$ \calK $ function such that $ \alpha_{1}(-x)\leq-\alpha_{2}(x) $ for all $ x\in[0,A] $ and some finite $ A>0 $. Then, there exists an extended class~$ \calK_{e} $ function $ \beta $ such that for all $ x_{1}\in[-A,\infty) $, $ x_{2}\in[0,A] $, it holds
	\begin{align}
		\label{0008_eq:lemma:upper bound class K}
		\alpha_{1}(x_{1}) + \alpha_{2}(x_{2}) \leq \beta(x_{1}+x_{2}).
	\end{align}
\end{lemma} 

The proof of the lemma in~\cite{Wiltz2025d} is constructive and provides an explicit derivation of~$ \beta $ in terms of $ \alpha_{1} $ and $ \alpha_{2} $.

%%%%%%%%%%%%%%%%%%%%%%%%%%%%%%%%%%%%%%%%%%%%%%%%%%%%%%%%%%%%%%%%%%%%%%%%%%%%%%%%
% MAIN RESULTS

\section{Main Results}

Let a time-invariant shiftable CBF $ b $ be given as well as a diffeomorphism~$ D $, under which dynamics~\eqref{0008_eq:dynamics} are equivariant. Based on these premises, we establish sufficient conditions such that 
\begin{align}
	\label{0008_eq:tv cbf}
	B(t,x)\coloneq b\big(D(x;\bm{p}(t))\big) + \bm{\lambda}(t) 
\end{align}
constitutes a time-varying CBF, where $ \bm{p}: \bbR_{\geq0} \rightarrow\calP $ varies the parameter of $ D $ over time, and $ \bm{\lambda}: \bbR_{\geq0}\rightarrow[0,\Lambda] $ is a time-varying scalar. This gives rise to the decoupled design of the state- and time-varying components. 

\subsection{Time-Variations via Equivariances}
\label{0008_subsec:time-variations via equivariances}

As by Definition~\ref{0008_def:equiv}, equivariances allow to infer a system’s behavior on a larger set of states from its behavior in a given state based on the defining diffeomorphism~$ D $. Analogously, if a CBF is known for an equivariant system, further CBFs can be obtained through the application of diffeomorphism $ D $.

\begin{proposition}
	Let a CBF $ b $, as defined in Definition~\ref{0008_def:cbf}, be given, and suppose dynamics~\eqref{0008_eq:dynamics} are equivariant with respect to a diffeomorphism $ D(\cdot;p) $, $ p\in\calP $. Then, $ b_{p}(x)\coloneq b(D(x;p)) $ is a CBF for any $ p\in\calP $.
\end{proposition}
\begin{proof}
	Since $ b $ is a CBF, it holds $ db(x;f(x,u)) \geq -\alpha(b(x)) $ by Definition~\ref{0008_def:cbf}. For $ b_{\sigma}(x) $, we derive
	\begin{align*}
		db_{\sigma}(x;f(x,u)) &= db(D(x;\sigma);\frac{\partial D}{\partial x}(x;\sigma) \, f(x,u)) \\
		&\stackrel{\eqref{0008_eq:def equivariance}}{=} db(D(x;\sigma);f(D(x;\sigma),\tilde{u})) \\
		&\stackrel{\eqref{0008_eq:cbf condition}}{\geq} -\alpha(b(D(x;\sigma))) = -\alpha(b_{\sigma}(x)),
	\end{align*}
	where we used the equivariance of dynamics~\eqref{0008_eq:dynamics} in terms of~\eqref{0008_eq:def equivariance}, and where $ \tilde{u}\!=\!D_{u}(u;\sigma) $ for some isomorphism~$ D_{u} $. As $ \tilde{u}\!\in\! D_{u}(\calU;\sigma)\!\subseteq\!\calU $ for any $ \sigma\in\calP $ due to the equivariance of~\eqref{0008_eq:dynamics} (see Definition~\ref{0008_def:equiv}), it follows that $ \sup_{u\in\calU}\{ db_{\sigma}(x;f(x,u)) \} \geq -\alpha(b_{\sigma}(x)) $. Thus, $ b_{\sigma} $ is a CBF. 
\end{proof}

For generalizing the result to time-varying parameters, we need to resort to CBFs defined on domains larger than their zero superlevel set~$ \calC $, which we termed shiftable CBFs. While gradient condition~\eqref{0008_eq:cbf condition} only guarantees on $ \calC $ that a maximum descend rate is not exceeded, the condition guarantees on the domain outside of $ \calC $, that is $ \calD\setminus\calC $, a certain minimum ascend rate. Based on this property, a condition on the time-variation in the parameter of diffeomorphism~$ \calD $ can be derived.

\begin{theorem}
	\label{0008_thm:equiv shift CBF}
	Let $ b: \bbR^{n}\rightarrow\bbR $ be a $ \Lambda $-shiftable CBF with respect to dynamics~\eqref{0008_eq:dynamics} with input constraint $ u\in\calU $ and with $ \alpha $ being its associated extended class~$ \calK_{e} $ function as per Definition~\ref{0008_def:shiftable cbf}. Moreover, let dynamics~\eqref{0008_eq:dynamics} be equivariant with respect to a diffeomorphism~$ D(\cdot;p): \bbR^{n} \rightarrow \bbR^{n} $, $ p\in\calP $, differentiable with respect to parameter~$ p $. Let $ \alpha_{p} $ be a convex or concave class~$ \calK $ function satisfying $ \alpha(-\xi) \leq -\alpha_{p}(\xi) $ for all $ \xi\in[0,\Lambda] $, and let $ \bm{p}:\bbR_{\geq 0} \rightarrow \calP $ be locally Lipschitz continuous. Then, for any $ \lambda\in[0,\Lambda] $,
	\begin{align}
		\label{0008_eq:equiv tv cbf simple}
		B(t,x) \coloneq b\big(D(x;\bm{p}(t))\big) + \lambda
	\end{align}
	is a time-varying CBF with respect to~\eqref{0008_eq:dynamics} with input constraint $ u\in\calU $, if 
	\begin{align}
		\label{0008_eq:p rate condition 1}
		||d\bm{p}(t;1)|| \leq \frac{c_{\alpha}}{\ell_{b}\ell_{D}} \qquad \forall t\geq 0,
	\end{align}
	where $ c_{\alpha} \coloneq \alpha_{p}(\lambda) $, and $ \ell_{b} $ is the local Lipschitz constant of $ b $ and $ \ell_{D} $ that of $ D $ with respect to its second argument. 
\end{theorem} 
\begin{proof}
	We start by noting that the Dini-derivative $ db(x;v+w) $ with $ v,w\in\bbR^{n} $ can be lower bounded as
	\begin{align}
		\label{0008_eq:thm:equiv shift aux 0}
		db(x;v+w) \geq db(x;v) - \ell_{b} ||w||,
	\end{align}
	where $ \ell_{b} $ denotes the local Lipschitz constant of $ b $, since
	\begin{align*}
		&db(x;v+w) = \liminf_{\varepsilon\downarrow0}\frac{b(x+\varepsilon(v+w)) - b(x)}{\varepsilon} \\
		&\quad= \liminf_{\varepsilon\downarrow 0}\frac{b(x+\varepsilon v) - b(x)}{\varepsilon}\\
		&\qquad+\liminf_{\varepsilon\downarrow0}\frac{b(x+\varepsilon(v+w)) - b(x+\varepsilon v)}{\varepsilon} \\
		&\quad\geq db(x;v) - \liminf_{\varepsilon\downarrow0}\frac{||b(x+\varepsilon(v+w)) - b(x+\varepsilon v)||}{\varepsilon} \\
		&\quad\geq db(x;v) - \liminf_{\varepsilon\downarrow0}\frac{\varepsilon\ell_{b} ||w||}{\varepsilon} = db(x;v) - \ell_{b} ||w||.
	\end{align*}
	For the application of the Dini derivative to a composition of functions, we recall that for locally Lipschitz continuous functions $ q $, $ r $ defined on suitable spaces and their composition $ s = q\circ r $, it holds
	\begin{subequations}
		\label{0008_eq:chain rule dini derivative}
		\begin{align}
			ds(x;v) &= \liminf_{\varepsilon\downarrow 0} \frac{q(r(x+\varepsilon v)) - q(r(x))}{\varepsilon} \\
			&= \liminf_{\tilde{\varepsilon}\downarrow0} \frac{q(r(x)+\tilde{\varepsilon}\, dr(x;v) + \calO(\tilde{\varepsilon}^{2})) - q(r(x))}{\tilde{\varepsilon}} \\
			&=dq(r(x);dr(x;v)).
		\end{align}
	\end{subequations}
	Equipped with this, we find that, for $ u\in\calU $, it holds
	\begin{subequations}
		\label{0008_eq:thm:equiv shift aux 1}
		\begin{align}
			\label{0008_seq:thm:equiv shift aux 1.1}
			&dB(t,x;1,f(x,u)) \\
			\label{0008_seq:thm:equiv shift aux 1.2}
			&\quad\stackrel{\eqref{0008_eq:chain rule dini derivative}}{=} db\Big(D(x;\bm{p}(t));dD\big(x,\bm{p}(t); f(x,u),d\bm{p}(t;1)\big)\Big) \\
			\label{0008_seq:thm:equiv shift aux 1.3}
			&\quad= db\Big(D(x;\bm{p}(t));\frac{\partial D}{\partial x} \, f(x,u) + \frac{\partial D}{\partial p} \, d\bm{p}(t;1) \Big) \\
			\label{0008_seq:thm:equiv shift aux 1.4}
			&\quad\stackrel{\eqref{0008_eq:thm:equiv shift aux 0}}{\geq} db\Big(D(x;\bm{p}(t));\frac{\partial D}{\partial x} \, f(x,u)\Big)- \ell_{b}\,\bigg|\bigg|\frac{\partial D}{\partial p} \, d\bm{p}(t;1)\bigg|\bigg|,
		\end{align}
	\end{subequations}
	where the differentiability of $ D $ is applied in~\eqref{0008_seq:thm:equiv shift aux 1.3}; the argument $ (x;p(t)) $ of $ \tfrac{\partial D}{\partial x} $ and $ \tfrac{\partial D}{\partial p} $ is omitted for brevity. Employing the equivariance of~\eqref{0008_eq:dynamics} as by~\eqref{0008_eq:def equivariance}, where we identify $ J_{D} $ with $ \tfrac{\partial D}{\partial x} $, we upper-bound the first summand in~\eqref{0008_seq:thm:equiv shift aux 1.4} as
	\begin{subequations}
		\label{0008_eq:thm:equiv shift aux 2}
		\begin{align}
			&db\Big(D(x;\bm{p}(t));J_{D}(x;\bm{p}(t)) \, f(x,u)\Big) \\
			&\qquad\stackrel{\eqref{0008_eq:def equivariance}}{=} db\Big(D(x;\bm{p}(t)); f(D(x;\bm{p}(t)),\tilde{u})\Big) \\ 
			&\qquad\stackrel{\eqref{0008_eq:cbf condition}}{\geq} -\alpha(b(D(x;\bm{p}(t)))) 
		\end{align}
	\end{subequations}
	with $ \tilde{u}\coloneq D_{u}(u;\bm{p}(t)) $, which lies in $ \calU $ as by Definition~\ref{0008_def:equiv}. Utilizing that, due to the Lipschitz continuity of $ b $ and $ D $ in its second argument, it holds $ |db(\tilde{x};\cdot)|\leq\ell_{b} $ at any $ \tilde{x}\in\bbR^{n} $ and $ || \tfrac{\partial D}{\partial p} || \leq \ell_{D} $, we obtain for the second summand in~\eqref{0008_seq:thm:equiv shift aux 1.4} that
	\begin{align}
		\label{0008_eq:thm:equiv shift aux 3}
		\ell_{b}\,\bigg|\bigg|\frac{\partial D}{\partial p} \, d\bm{p}(t;1)\bigg|\bigg| \leq \ell_{b} \ell_{D} ||d\bm{p}(t;1)|| \stackrel{\eqref{0008_eq:p rate condition 1}}{\leq} \alpha_{p}(\lambda).
	\end{align}
	Summarizing the results from \eqref{0008_eq:thm:equiv shift aux 1}-\eqref{0008_eq:thm:equiv shift aux 3}, we obtain by additionally employing Lemma~\ref{0008_lemma:upper bound class K function}
	\begin{subequations}
		\label{0008_eq:thm:equiv shift aux 4}
		\begin{align}
			dB(t,x;1,f(x,u)) &\geq -\alpha\big(b(D(x;\bm{p}(t)))\big) - \alpha_{p}\big(\lambda\big) \\
			&\stackrel{\eqref{0008_eq:lemma:upper bound class K}}{\geq} -\beta\big(b(D(x;\bm{p}(t))) + \lambda\big),
		\end{align}
	\end{subequations}
	where $ \beta $ is an extended class~$ \calK_{e} $ function. Noting that $ B $ is also inherently locally Lipschitz continuous, we conclude that $ B $ is a time-varying CBF.
\end{proof}

The result allows to decouple the design of a time-varying CBF into a time-invariant component, namely the shiftable CBF~$ b $, and a time-varying component represented by diffeomorphism $ D $ and its time-varying parameter $ \bm{p} $. 

Particularly, the theorem establishes bounds within which the time-varying parameter~$ \bm{p} $ can be freely chosen, while the system characteristics that restrict the admissible rates of change are encoded in the constants $ c_{\alpha} $, $ \ell_{b} $ and $ \ell_{D} $, giving rise to the decoupling nature of the result. The implications of these three constants for applying the theorem are as follows: whereas $ c_{\alpha} $ is fundamental for characterizing the rate at which the CBF may vary, the Lipschitz constants $ \ell_{D} $ and $ \ell_{b} $ act as normalizing factors. All constants can be determined via established methods. Constant $ c_{\alpha} $ is directly derived from~$ \alpha $, which in turn can be obtained as an analytic function from data-driven or predictive synthesis methods along with the CBF~\cite{Lindemann2024a, Chen2024c, Wang2024c, Wiltz2025b}; a survey of suitable methods, including also analytical ones, can be found in~\cite[Sec.~V]{Wiltz2025d}. As transformation~$ D $ is commonly analytically derived and always differentiable, the derivation of $ \ell_{D} $ is straightforward. If $ b $ is also analytically known, the same methods as before are applicable, as $ b $ is Lipschitz continuous. Otherwise, standard techniques such as sampling gradients over the region of interest or automatic differentiation yield good approximations of~$ \ell_{b} $. 

\begin{remark}
	\label{0008_remark:lb relaxation}
	In practice, a safety filter only needs to be implemented for states that lie close to the boundary of the safe set~$ \calC_{\lambda} $ in order to guarantee forward invariance; these states correspond to the $ \varepsilon $-neighborhood $ \partial\calC_{\lambda}^{\varepsilon} \coloneq \{x \,| \, |b(x) + \lambda|\leq \varepsilon\} $. All other states lie well in the interior of the safe set and do not require safety filtering. Consequently, it suffices for $ \ell_{b} $ to be a local Lipschitz constant of $ b $ on~$ \partial\calC_{\lambda}^{\varepsilon} $. 
\end{remark}

\begin{remark}
	Lipschitz constant $ \ell_{D} $ can be omitted, when choosing transformation $ D $ directly as a time-varying function. In this case, \eqref{0008_eq:p rate condition 1} reduces to
	\begin{align*}
		\bigg|\bigg|\frac{\partial D}{\partial t}(x;t)\bigg|\bigg| \leq \frac{c_{\alpha}}{\ell_{b}}. 
	\end{align*}
\end{remark}

\subsection{Incorporating Uniform Time-Variations}

In addition to the time variation of $ b $ via transformation $ D(x;\bm{p}(t)) $, we now also allow the offset $ \lambda $ in~\eqref{0008_eq:equiv tv cbf simple} to vary over time. By varying the offset, the size of the zero superlevel set of the function 
\begin{align}
	\label{0008_eq:uniformly tv cbf}
	\widehat{B}(t,x) \coloneq b(x) + \bm{\lambda}(t)
\end{align}
can be enlarged or reduced, corresponding to an increase or decrease in $ \bm{\lambda} $, respectively. Here, $ b $ is considered to be a $ \Lambda $-shiftable CBF, and $ \bm{\lambda}: \bbR_{\geq0}\rightarrow[0,\Lambda] $ is a time-varying function that satisfies the following continuity properties. 
\begin{assumption}
	\label{0008_ass:continuity lambda}
	Let $ \bm{\lambda} $ be piecewise differentiable, upper semi-continuous and let
	\begin{align*}
		\lim_{t\uparrow t_{0}} \bm{\lambda}(t) < \lim_{t\downarrow t_{0}} \bm{\lambda}(t) \qquad \forall t_{0}\in\Omega,
	\end{align*}
	where $ \Omega\coloneq\{t \, | \, \bm{\lambda}(\cdot) \text{ is discontinuous at time } t\} $.
\end{assumption}

The relaxed continuity properties of~$ \bm{\lambda} $ account for the continuity properties in the time argument required for time-varying CBFs as per Assumption~\ref{0008_ass:relaxed continuity for tv cbf}. Based on the results from the previous section, we now establish conditions under which~\eqref{0008_eq:tv cbf} with a time-varying offset also constitutes a time-varying CBF.

\begin{theorem}
	\label{0008_thm:equiv shift CBF plus}
	Let $ \Lambda $-shiftable CBF $ b $, diffeomorphism $ D $ and extended class~$ \calK_{e} $ function $ \alpha_{p} $ be the same as in Theorem~\ref{0008_thm:equiv shift CBF}. Moreover, let $ \bm{p}: \bbR_{\geq0}\rightarrow\calP $ be locally Lipschitz continuous and let $ \bm{\lambda}:\bbR_{\geq 0} \rightarrow [0,\Lambda] $ satisfy Assumption~\ref{0008_ass:continuity lambda}. If
	\begin{align}
		\label{0008_eq:p rate condition 2}
		\ell_{b}\ell_{D} ||d\bm{p}(t;1)|| - d\bm{\lambda}(t;1) \leq \alpha_{p}(\bm{\lambda}(t)) \qquad \forall t\geq 0, 
	\end{align}
	where $ \ell_{b} $ is the local Lipschitz constant of $ b $ on $ \calD $ and $ \ell_{D} $ the local Lipschitz constant of $ D $ with respect to its second argument, then $ B $ defined in~\eqref{0008_eq:tv cbf} as 
	\begin{align*}
		B(t,x)\coloneq b\big(D(x;\bm{p}(t))\big) + \bm{\lambda}(t) 
	\end{align*}
	is a time-varying CBF with respect to~\eqref{0008_eq:dynamics} with input constraint $ u\in\calU $.
\end{theorem}
\begin{proof}
	Based on intermediate results from the proof of Theorem~\ref{0008_thm:equiv shift CBF}, we derive
	\begin{align*}
		&dB(t,x;1,f(x,u)) \stackrel{\eqref{0008_eq:thm:equiv shift aux 1}}{\geq} db\Big(D(x;\bm{p}(t));\frac{\partial D}{\partial x} f(x,u)\Big) \\
		&\qquad\quad - \ell_{b} \bigg|\bigg|\frac{\partial D}{\partial p} d\bm{p}(t;1)\bigg|\bigg| + d\bm{\lambda}(t;1) \\
		&\quad\stackrel{\eqref{0008_eq:thm:equiv shift aux 2},\eqref{0008_eq:p rate condition 2}}{\geq} -\alpha\Big(b\big(D(x;\bm{p}(t))\big)\Big) - \alpha_{p}(\bm{\lambda}(t)) \\
		&\qquad \stackrel{\eqref{0008_eq:lemma:upper bound class K}}{\geq} -\beta\Big(b\big(D(x;\bm{p}(t))\big)+\bm{\lambda}(t)\Big),
	\end{align*}
	which corresponds to~\eqref{0008_eq:time-varying cbf condition}, and the result follows.
\end{proof}

The requirements on the Lipschitz constants can be relaxed analogously to before. In particular, if diffeomorphism $ D $ is directly parameterized over the time-domain, then~\eqref{0008_eq:p rate condition 2} reduces to
\begin{align*}
	\ell_{b} \bigg|\bigg|\frac{\partial D}{\partial t}(x;t)\bigg|\bigg| - d\bm{\lambda}(t;1) \leq \alpha_{p}(\bm{\lambda}(t)).
\end{align*}

\subsection{Time-Varying Constraints}

Before providing guidelines on the practical design of time-varying CBFs through equivariances, we briefly discuss classes of time-varying constraints that can be handled with our method. As such, the proposed approach allows for the guaranteed satisfaction of time-varying constraints beyond uniformly time-varying ones. Specifically, for systems that are equivariant with respect to $ D(\cdot;p) $, $ p\in\calP $, any time-varying constraint $ x(t)\in\calH(t)\coloneq\{x \, | \, h(t,x)\geq0\} $, where $ h:\bbR_{\geq0}\times\bbR^{n}\rightarrow\bbR $ takes the form 
\begin{align}
	\label{0008_eq:constraint function equiv tv}
	h(t,x) \coloneq \bar{h}\big(D(x;\bm{q}(t))\big) + \bm{\gamma}(t),
\end{align}
can be satisfied. Here, $ \bar{h}:\bbR^{n}\rightarrow\bbR $ is a locally Lipschitz continuous function, defining the underlying time-invariant constraint; the time-varying parameter $ \bm{q}: \bbR_{\geq 0}\rightarrow\calP $ is also locally Lipschitz continuous; and the time-varying offset $ \bm{\gamma}:\bbR_{\geq0}\rightarrow\bbR^{n} $ is assumed to satisfy Assumption~\ref{0008_ass:continuity lambda}. Whereas $ \bm{\gamma} $ uniformly varies the time-invariant constraint function~$ \bar{h} $ in magnitude and thereby shrinking or increasing the set of feasible states, the type of time-variations induced through $ \bm{q} $ depends on the equivariances that the dynamics allow for. Examples on the derivation of equivariances and the corresponding diffeomorphisms can be found in~\cite{Wiltz2025c}. Especially kinematics induce equivariances into dynamics~\cite{Apraez2025}. As such, mobile robots kinematically modeled as for instance as bicycle or unicycles are equivariant with respect to translations and rotations, and desired translational or rotational movements of the constraints can be specified in terms of $ \bm{q} $ in this context. 

We note that clearly also any constraint given through a general constraint function $ h':\bbR_{\geq 0}\times\bbR^{n}\rightarrow\bbR $ can be handled if there exists an under approximation in terms of $ h $ as in~\eqref{0008_eq:constraint function equiv tv} such that $ h'(t,x)\leq h(t,x) $ for any $ t $ and $ x $. The satisfaction of the more general constraint then follows, though more conservatively, from the satisfaction of $ x\in\calH(t) $ for $ t\geq 0 $. 

\subsection{Decoupled Design}

The decoupled design procedure for a time-varying CBF is summarized as follows: first, determine a time-invariant CBF~$ b $ and its associated extended class~$ \calK_{e} $ function on a domain $ \calD\subseteq\bbR^{n} $ such that $ b(x)\leq\bar{h}(x) $; we commented on this design step before in Section~\ref{0008_subsec:time-variations via equivariances} and a survey of available methods is found in~\cite[Sec.~V]{Wiltz2025d}. Let the largest zero-super level set contained in $ \calD $ be $ \calC_{\Lambda} \coloneq \{ x \, | \, b(x)\geq -\Lambda \} $, $ \Lambda > 0 $. Based on this, the second step introduces time-variations via the design of $ \bm{p} $ and $ \bm{\lambda} $. In the remainder of the section, we provide some guidance on their choice.

Given constraint~\eqref{0008_eq:constraint function equiv tv}, choose $ \bm{\lambda}(t)\leq\bm{\gamma}(t) $ for all $ t\geq 0 $, and verify that condition~\eqref{0008_eq:p rate condition 1} or~\eqref{0008_eq:p rate condition 2} is satisfied for constant or time-varying~$ \bm{\lambda} $, respectively. Then, $ B $ in~\eqref{0008_eq:tv cbf} encodes constraint~\eqref{0008_eq:constraint function equiv tv}, and its satisfaction is ensured through the forward invariance of the zero superlevel set of $ B $. If the prior condition is not satisfied, a feasible approximation  can be obtained by rescaling time. For instance, if $ \bm{\lambda} $ is constant, $ t $ can be substituted with a strictly increasing function $ \tau:\bbR_{\geq0}\rightarrow\bbR_{\geq 0} $ satisfying
\begin{align*}
	\frac{\partial\tau}{\partial t}(t) \leq \frac{c_{\alpha}}{\ell_{b}\ell_{D}} \frac{1}{d\bm{p}(\tau(t);1)}.
\end{align*}
Then, \eqref{0008_eq:p rate condition 1} holds for $ \bm{p}\circ\tau $. The case for time-varying $ \bm{\lambda} $ is analogous. 

If the constraint is not specified a priori, we have more flexibility and can adapt to the system's dynamic capabilities. To this end, replace first $ \bm{\lambda} $ in~\eqref{0008_eq:p rate condition 2} by a constant $ \bar{\lambda} $ and set $ d\bm{\lambda} $ to zero; let $ \bar{\lambda} $ such that $ b(x(0))\geq -\bar{\lambda} $. Then, \eqref{0008_eq:p rate condition 2} reduces to~\eqref{0008_eq:p rate condition 1}, and any $ \bm{p} $ satisfying the condition can be chosen. Subsequently, substitute the chosen $ \bm{p} $ into~\eqref{0008_eq:p rate condition 2}, which then becomes a differential inequality in $ \bm{\lambda} $. If $ \bm{\lambda} $ is constant, one can directly start with~\eqref{0008_eq:p rate condition 1} and the method becomes straightforward. Alternatively, if the design of a time-varying~$ \bm{\lambda} $ is prioritized, select first $ \bm{p}\equiv 0 $ and choose $ \bm{\lambda} $ such that 
\begin{align*}
	-d\bm{\lambda}(t;1) \leq \alpha_{p}(\bm{\lambda}(t)),
\end{align*}
for instance, by solving a differential equation. Subsequently, insert the selected $ \bm{\lambda} $ into~\eqref{0008_eq:p rate condition 2} and choose $ \bm{p} $ such that 
\begin{align*}
	||d\bm{p}(t;1)|| \leq \frac{\alpha_{p}(\bm{\lambda}(t)) + d\bm{\lambda}(t;1)}{\ell_b\ell_D}.
\end{align*}
The right-hand side can be viewed as the margin left for selecting~$ \bm{p} $ after the design of~$ \bm{\lambda} $.

%%%%%%%%%%%%%%%%%%%%%%%%%%%%%%%%%%%%%%%%%%%%%%%%%%%%%%%%%%%%%%%%%%%%%%%%%%%%%%%%
% SIMULATIONS

\section{Numerical Examples}

We demonstrate the approach in two scenarios. At first, a compact forward control-invariant set is shifted along waypoints by varying parameter $ \bm{p} $ for trajectory tracking. Secondly, two systems with degenerate controllability properties avoid collisions with dynamically evolving obstacles whose variations are a priori unknown to the controllers. An animation of the simulation results is available on Youtube\footnote{\url{https://www.youtube.com/watch?v=1f_e7tTwWNI}}.

\subsection{Waypoint Tracking via Forward Invariant Sets}

\begin{figure}[t]
	\centering
	\includegraphics[width=0.65\columnwidth]{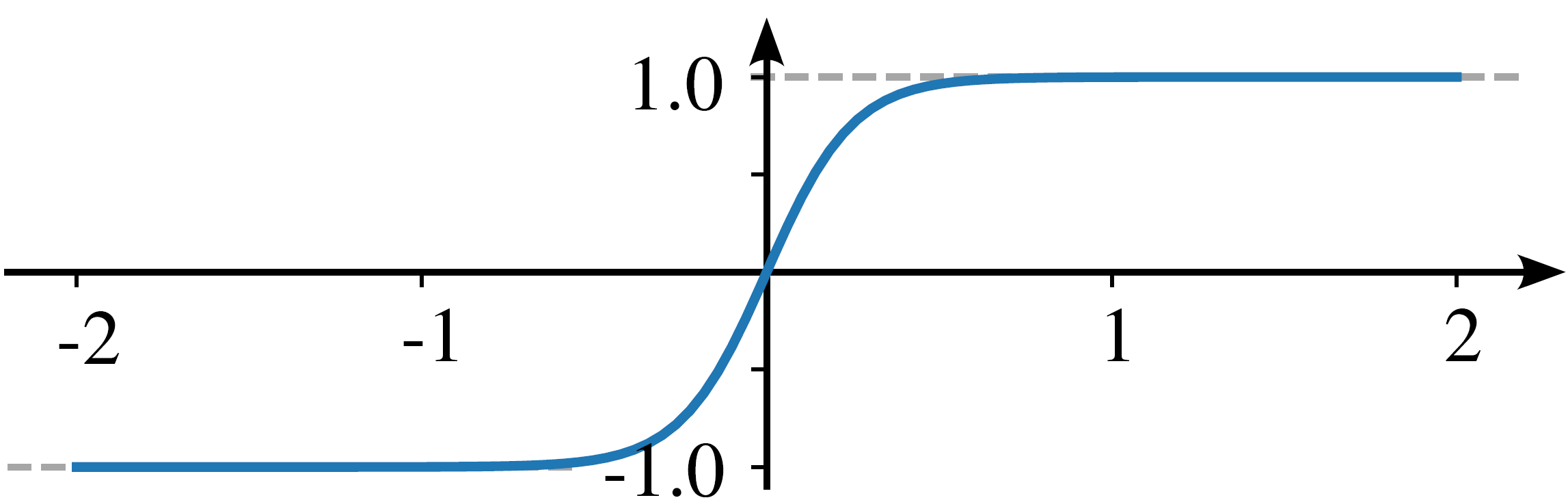}
	\caption{Sigmoid function $ \sigma $ for $ c_{1}=1 $, $ c_{2}=8 $.}
	\vspace{-\baselineskip}
	\label{0008_fig:sigmoid_like}
\end{figure}

We start by illustrating the design principle with a single and a double integrator, both systems with favorable controllability properties, given as $ \dot{x} = u $ and $ \ddot{x}=u $. For brevity, define the sigmoid like function
\begin{align*}
	\sigma(\xi;c_1,c_2)\coloneq 2 c_{1} \left(\frac{1}{1+e^{-c_{2}\xi}} - 0.5\right)
\end{align*}
with parameters $ c_{1}, c_{2} \in\bbR $, see Figure~\ref{0008_fig:sigmoid_like}. A shiftable CBF for the single integrator is $ b_{\text{SI}}(x) = -||x|| $ with associated class~$ \calK_{e} $ function $ \alpha_{\text{SI}}(\xi)\coloneq\sigma(\xi;1,8) $ and input constraint $ u\in\calU_{\text{SI}}=[-1,1]^{n} $. It holds
\begin{align}
	\label{0008_eq:exmp waypoint tacking eq 1}
	\frac{db_{\text{SI}}}{dt}(x) = -\frac{x^{T}}{||x||} \dot{x} \geq -\alpha_{\text{SI}}(b_{\text{SI}}(x))
\end{align}
with $ \dot{x} \!=\! u $, $ u\!\in\!\calU_{\text{SI}} $. Evidently, the single integrator is equi\-variant with respect to translations $ D_{\text{SI}}(x;p) = x-p $. Based on this, we design a time-varying CBF~$ B_{\text{SI}} $ of the form~\eqref{0008_eq:equiv tv cbf simple} to track a series of waypoints by shifting a forward control invariant circle of radius $ r_{1}=0.25 $. To this end, set to $ \lambda= r_{1} $, and choose~$ \bm{p} $ as trajectory connecting the waypoints such that 
\begin{align}
	\label{0008_eq:exmp waypoint tacking eq 2}
	||d\bm{p}(t;1)||\leq c_{\alpha_{\text{SI}}} \approx 0.76,
\end{align}
ensuring~\eqref{0008_eq:p rate condition 1}. Here, the local Lipschitz constants are $ \ell_{b_{\text{SI}}} =\ell_{D} = 1 $. By Theorem~\ref{0008_thm:equiv shift CBF}, $ B_{\text{SI}} $ is a time-varying CBF.

\begin{figure}
	\begin{subfigure}[b]{1.0\linewidth}
		\centering
		\includegraphics[width=0.65\columnwidth]{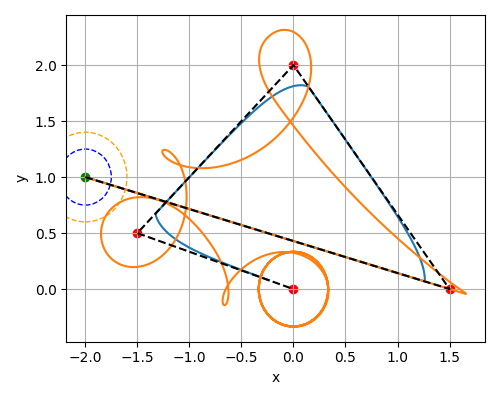}
		\vspace{-0.5\baselineskip}
		\caption{Trajectories.}
		\label{0008_sub_fig:integrator_sys_traj}
	\end{subfigure}
	\linebreak
	\begin{subfigure}[b]{0.48\linewidth}
		\centering
		\includegraphics[width=1.0\columnwidth]{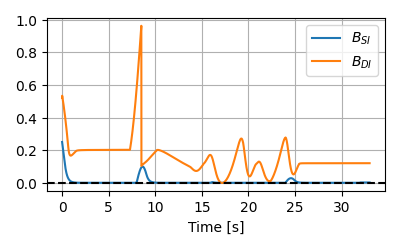}
		\caption{Values time-varying CBFs.}
		\label{0008_sub_fig:Integrator_sys_b_values}
	\end{subfigure}
	\hfill
	\begin{subfigure}[b]{0.51\linewidth}
		\centering
		\includegraphics[width=0.93\columnwidth]{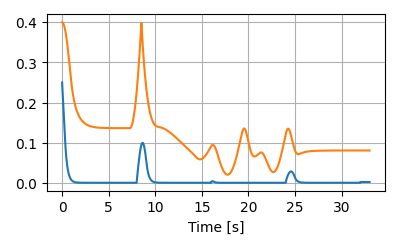}
		\caption{Distance to constraint boundary.}
		\label{0008_sub_fig:Integrator_sys_constraint_satisfaction}
	\end{subfigure}
	\caption{Waypoint tracking via invariance of translating radial constraint. Simulation results for single ({\color{blue}blue}) and double integrator ({\color{orange}orange}).}
	\label{0008_fig:integrator_sys}
\end{figure}

For the double integrator example, we illustrate the application of the proposed approach to high-order constraints. Observe that the double integrator is equivariant with respect to $ D_{\text{DI}}(x,\dot{x};p,\dot{p}) = \begin{bsmallmatrix}
	x - p \\ \dot{x} - \dot{p}
\end{bsmallmatrix} $. As before, consider a radial constraint moving along a series of waypoints, now with $ r_{2}=0.4 $. Noting that we now control acceleration rather than velocity, a backstepping approach is required to satisfy~\eqref{0008_eq:exmp waypoint tacking eq 1}. To this end, define auxiliary function $ b_{\text{DI}}(x,\dot{x}) = \frac{db_{\text{SI}}}{dt}(x) + \alpha_{\text{SI}}(b_{\text{SI}}(x)) $ for $ x\neq 0 $, whose zero superlevel set encodes~\eqref{0008_eq:exmp waypoint tacking eq 1}. Though discontinuous at the origin, it satisfies~\eqref{0008_eq:cbf condition} for $ ||x||\leq r_{2}\pm 0.1 $, that is, near the constraint boundaries, for extended~$ \calK_{e} $ function $ \alpha_{\text{DI}}(\xi) = \sigma(\xi;3,6) $ and $ u\in\calU_{\text{DI}}=[-7.5,7.5]^{n} $. As by Remark~\ref{0008_remark:lb relaxation}, Theorem~\ref{0008_thm:equiv shift CBF} is then still applicable. Similarly to before, we construct a time-varying function~$ B_{\text{DI}} $ of the form~\eqref{0008_eq:equiv tv cbf simple} by choosing $ \lambda = r_{2} $, and $ \bm{p} $, $ \dot{\bm{p}} $ such that
\begin{align}
	\label{0008_eq:exmp waypoint tacking eq 3}
	\left|\left|\begin{bsmallmatrix}
		\dot{\bm{p}}(t) \\ \ddot{\bm{p}}(t)
	\end{bsmallmatrix} \right|\right| \leq ||\dot{\bm{p}}(t)|| + ||\ddot{\bm{p}}(t)|| \leq \frac{c_{\alpha_{\text{DI}}}}{\ell_{b_{\text{DI}}} \ell_{D}} \approx 0.75,
\end{align}
with $ c_{\alpha_{\text{DI}}} = \alpha_{\text{DI}}(r_{2}) \approx 3.33 $, $ \ell_{b_{\text{DI}}} \approx 4.4 $ and $ \ell_{D} = 1 $. Here, $ \ell_{b_{\text{DI}}} $ is the local Lipschitz constant of $ b_{\text{DI}} $ for $ ||x||\leq r_{2}\pm 0.1 $. Additionally, $ \bm{p} $ needs to satisfy~\eqref{0008_eq:exmp waypoint tacking eq 2}, here however directly implied by the stricter condition~\eqref{0008_eq:exmp waypoint tacking eq 3}. The additional compatible CBF $ b_{\text{vel}}(x,\dot{x}) = -||\dot{x}|| + 1 $ accounts for velocity constraint~$ \dot{x}\in[-1,1]^{2} $. 

The controller is for both systems a standard quadratic program $ u^{\ast} = \argmin_{u\in\calU} || u - u_{\text{ref}} || $ subject to the respective CBF-based gradient conditions. Simulation results for~$ u_{\text{ref}} \equiv 0 $ are shown in Figure~\ref{0008_fig:integrator_sys}. Clearly, the state constraints are satisfied at all times despite the decoupled design.

\subsection{Collision Avoidance for Dynamically Evolving Obstacles} 

\begin{figure}
	\begin{subfigure}[b]{1.0\linewidth}
		\centering
		\includegraphics[width=1.0\columnwidth]{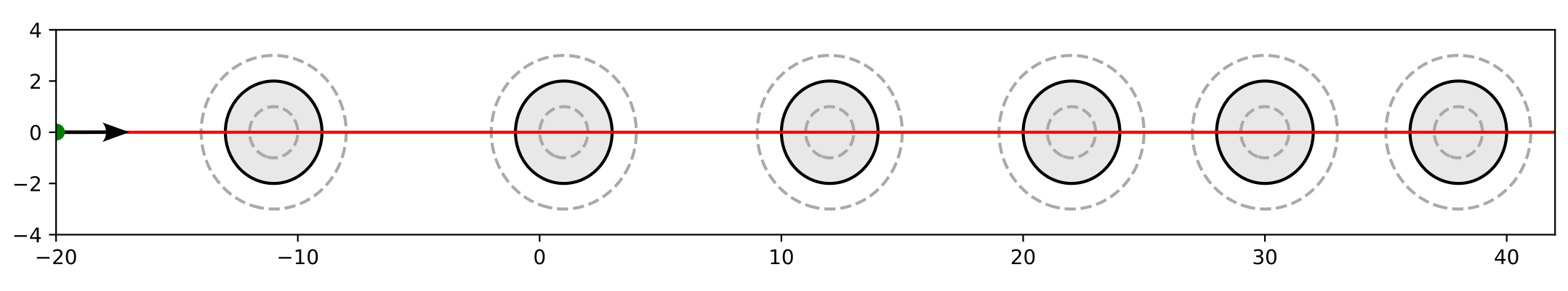}
		\caption{Obstacle configuration. Obstacles are moving with a constant speed horizontally towards the left while varying sinusoidally around the red line. In the first scenario, the radius is constant ($ r_{2} = 2 $, solid circle), in the second time-varying ($ r_{2}(t)\in[1,3] $, dashed circles). The systems, starting at initial position $ (0,-20) $, move towards the right tracking the red reference.}
		\label{0008_sub_fig:moving_obstacles_visualization}
	\end{subfigure}
	\linebreak
	\begin{subfigure}[b]{1.0\linewidth}
		\centering
		\includegraphics[width=0.9\columnwidth]{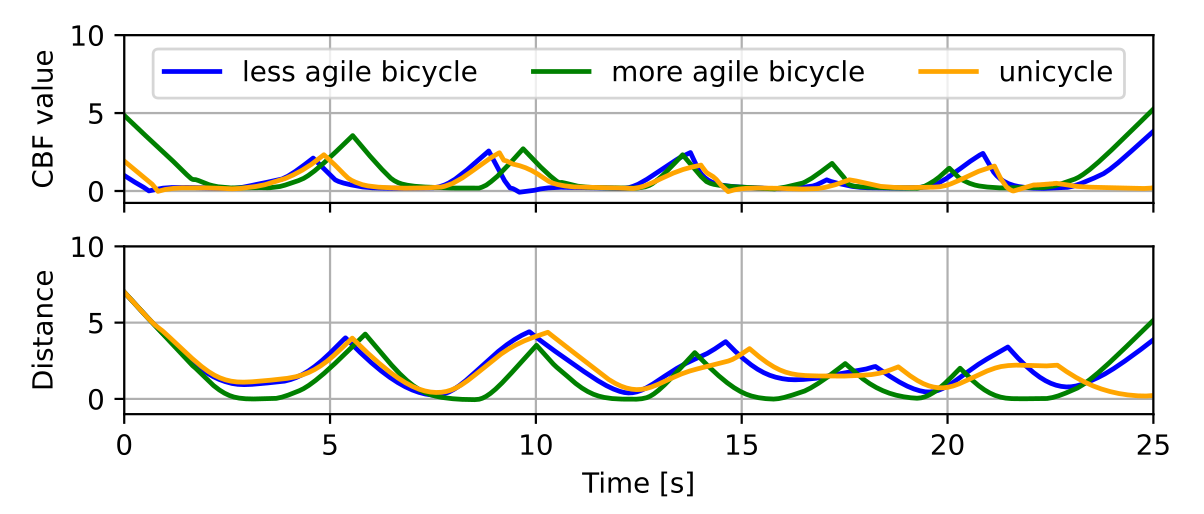}
		\caption{Simulation results: constant obstacle radius.}
		\label{0008_sub_fig:moving_obstacles_radius_static}
	\end{subfigure}
	\linebreak
	\begin{subfigure}[b]{1.0\linewidth}
		\centering
		\includegraphics[width=0.9\columnwidth]{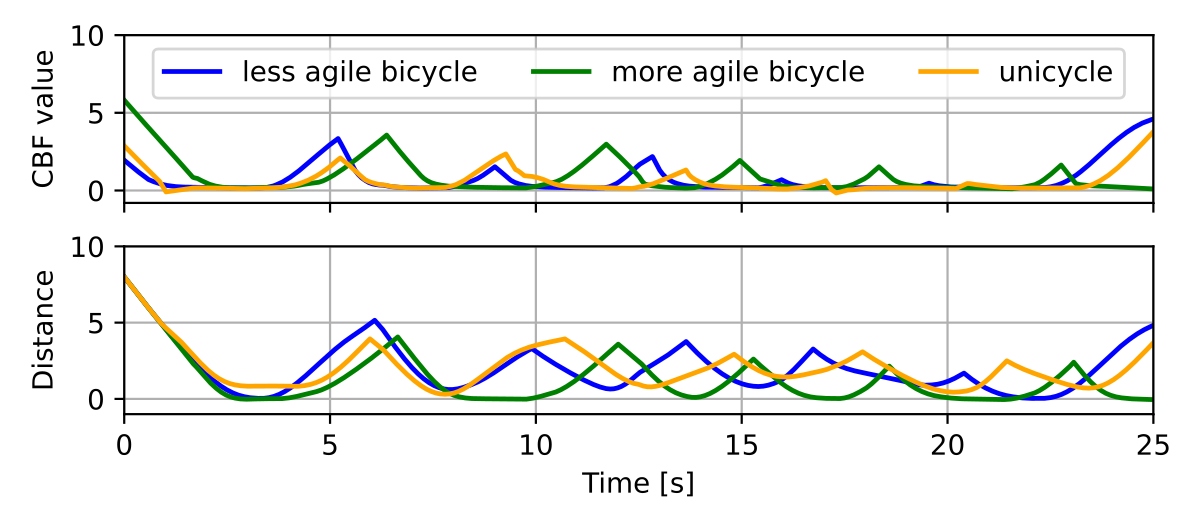}
		\caption{Simulation results: time-varying obstacle radius.}
		\label{0008_sub_fig:moving_obstacles_radius_tv}
	\end{subfigure}
	\caption{Avoidance of dynamically evolving obstacles through bicycle and unicycle. (a) Problem setting and (b)-(c) distance to obstacles and values of time-varying CBFs.}
	\label{0008_fig:moving_obstacles}
\end{figure}

We now turn towards the avoidance of dynamically evolving obstacles through systems with degenerate controllability properties. In particular, we consider two input constrained bicycles with dynamics $ \dot{x} = v \cos(\psi+\beta(\zeta))$ ,  $ \dot{y} = v\sin(\psi+\beta(\zeta))$,  $ \dot{\psi}=\frac{1}{L}v\cos(\beta(\zeta))\tan(\zeta) $, where the velocity $ v $ and the steering angle are control inputs, subject to input constraints $ (v,\zeta)\in\calU_{\text{B1}}\coloneq [1,2]\times[-\tfrac{20\pi}{180},\tfrac{20\pi}{180}] $ and $ (v,\zeta)\in\calU_{\text{B2}}\coloneq [1,2]\times[-\tfrac{45\pi}{180},\tfrac{45\pi}{180}] $ rendering the first bicycle less agile and the second more. Furthermore, we consider an input constrained unicycle with dynamics $ \dot{x}=v\cos(\psi) $, $ \dot{y}=v\sin(\psi) $, $ \dot{\psi}=\omega $, where velocity $ v $ and rotation rate $ \omega $ are control inputs, subject to input constraint $ (v,\omega)\in\calU_{\text{U}}\coloneq [1,2]\times[-0.9,0.9] $. The control objective is to track a reference while avoiding obstacles that vary over time and whose distances decrease, as illustrated in Figure~\ref{0008_sub_fig:moving_obstacles_visualization}. While the systems move to the right, the obstacles travel left at a constant speed and oscillate vertically in a sinusoidal pattern around the reference. In a first scenario, the obstacles have a fixed radius ($ r_{1}=2 $), whereas in the second, the radius varies over time ($ r_{2}(t)\in[1,3] $). 

Noting that underactuation and input constraints render the CBF synthesis problem non-trivial, we follow~\cite{Wiltz2025b} to compute a shiftable CBF and its associated extended class~$ \calK_{e} $ function numerically. The dynamics under consideration are translationally equivariant with respect to $ D(\bm{x};p)\coloneq \bm{x}-\begin{bsmallmatrix} p \\ 0 \end{bsmallmatrix} $ with $ \bm{x}=[x,y,\psi]^{T} $ and $ p\in\bbR^{2} $. In the first scenario, the time-varying CBFs are derived based on Theorem~\ref{0008_thm:equiv shift CBF} with $ \lambda = r_{1} $. The time-variation of the obstacles is a translation given through $ \bm{p}:\bbR_{\geq 0}\rightarrow\bbR^{2} $, satisfying~\eqref{0008_eq:p rate condition 1} with $ c_{\alpha} \geq 1.99 $, $ \ell_{b} \approx 1.0 $ and $ \ell_{D} = 1.0 $ (the Lipschitz constants are determined for fixed orientations as the orientation is invariant under $ D $). In the second scenario with time-varying radius, we employ Theorem~\ref{0008_thm:equiv shift CBF plus} instead. For the obstacles, consider the same $ \bm{p} $ as before, and choose $ \bm{\lambda}(t) \equiv r_{2}(t) $. By~\eqref{0008_eq:p rate condition 2} and $ ||d\bm{p}(t;1)||\leq c_{\alpha} $, we select~$ \bm{\lambda} $ such that 
\begin{align*}
	d\bm{\lambda}(t;1) \geq \ell_{b}\ell_{D} c_{\alpha} - \alpha_{p}(\bm{\lambda}(t)) \geq 1.99 - \alpha_{p}(1) \qquad \forall t\geq 0;
\end{align*}
in the last inequality, we used that $ r_{2}(t)=\bm{\lambda}(t)\geq 1 $ for all $ t\geq 0 $. By Theorem~\ref{0008_thm:equiv shift CBF} and~\ref{0008_thm:equiv shift CBF plus}, respectively, functions~\eqref{0008_eq:tv cbf} and~\eqref{0008_eq:equiv tv cbf simple} are time-varying CBFs. As before, the controller for each system is implemented as the standard QP, subject to the CBF gradient condition, with a tracking controller $u_{\text{ref}}$ for following the reference. The simulation results are shown in Figure~\ref{0008_fig:moving_obstacles}. We point out that neither $ \bm{p} $ nor $ \bm{\lambda} $ are a priori known to the controller. Instead, the controller only accesses the obstacle configuration at each time instant; the rate of change is estimated via a numerical first-order approximation based on the past and current configuration.

%%%%%%%%%%%%%%%%%%%%%%%%%%%%%%%%%%%%%%%%%%%%%%%%%%%%%%%%%%%%%%%%%%%%%%%%%%%%%%%%
% CONCLUSION

\section{Conclusion}

We presented a systematic and computationally efficient method for designing time-varying CBFs by decoupling the problem into the computation of a particular time-invariant CBF, and the design of time-dependent transformations based on equivariances and uniform shifts. This enables the generation of a broad class of time-varying CBFs without prior knowledge of the specific time variations. Thereby, our approach is applicable to uncertain environments that vary at bounded rates.

%%%%%%%%%%%%%%%%%%%%%%%%%%%%%%%%%%%%%%%%%%%%%%%%%%%%%%%%%%%%%%%%%%%%%%%%%%%%%%%%
%\hspace{-0.0cm}

\balance

\bibliographystyle{IEEEtran}
\bibliography{/Users/wiltz/CloudStation/JabBib/Research/000_MyLibrary}

\end{document}